\documentclass[fleqn,usenatbib]{mnras}

\usepackage{newtxtext,newtxmath}
\usepackage{amsmath}
\usepackage{mathtools}

\newcommand{\matr}[1]{\mathbf{#1}}
\usepackage[T1]{fontenc}

\DeclareRobustCommand{\VAN}[3]{#2}
\let\VANthebibliography\thebibliography
\def\thebibliography{\DeclareRobustCommand{\VAN}[3]{##3}\VANthebibliography}

\usepackage{graphicx}	% Including figure files
\usepackage{amsmath}	% Advanced maths commands
\usepackage{amssymb}	% Extra maths symbols

\title[PLD for \textit{Spitzer} Microlens Parallax Survey]{Pixel Level Decorrelation in Service of the \textit{Spitzer} Microlens Parallax Survey}

\author[L. Dang et al.]{
Lisa Dang,$^{1,2,3}$\thanks{E-mail: lisa.dang@physics.mcgill.ca}
S. Calchi Novati,$^{3}$
S. Carey$^{3}$
and N. B. Cowan$^{1,2,4}$
\\
$^{1}$Department of Physics, McGill University, 3600 University St, Montr\'eal, QC H3A 2T8, Canada\\
$^{2}$McGill Space Institute, Institute for Research on Exoplanets, Centre de Recherche en Astrophysique du Qu\'ebec, Technologies in Exo-Planetary Sciences\\
$^{3}$IPAC, Mail Code 100-22, Caltech, 1200 E. California Blvd., Pasadena, CA 91125, USA\\
$^{4}$Department of Earth and Planetary Science, McGill University, 3450 University St, Montr\'eal, QC H3A 2A7, Canada
}

\date{Accepted XXX. Received YYY; in original form ZZZ}

\pubyear{2020}

\begin{document}
\label{firstpage}
\pagerange{\pageref{firstpage}--\pageref{lastpage}}
\maketitle

\begin{abstract}
Microlens parallax measurements combining space-based and ground-based observatories can be used to study planetary demographics. In recent years, the Spitzer Space Telescope was used as a microlens parallax satellite. Meanwhile, \textit{Spitzer} IRAC has been employed to study short-period exoplanets and their atmospheres. As these investigations require exquisite photometry, they motivated the development of numerous self-calibration techniques now widely used in the exoplanet atmosphere community. Specifically, Pixel Level Decorrelation (PLD) was developed for starring-mode observations in uncrowded fields. We adapt and extend PLD to make it suitable for observations obtained as part of the \textit{Spitzer} Microlens Parallax Campaign. We apply our method to two previously published microlensing events, OGLE-2017-BLG-1140 and OGLE-2015-BLG-0448, and compare its performance to the state-of-the-art pipeline used to analyses \textit{Spitzer} microlensing observation. We find that our method yields photometry 1.5--6 times as precise as previously published. %\textbf{
In addition to being useful for \textit{Spitzer}, a similar approach could improve microlensing photometry with the Nancy Grace Roman Space Telescope.%}
\end{abstract}


%%%%%%%%%%%%%%%%%%%%%%%%%%%%%%%%%%%%%%%%%%%%%%%%%%

%%%%%%%%%%%%%%%%% BODY OF PAPER %%%%%%%%%%%%%%%%%%

\section{Introduction}

Gravitational microlensing is a powerful tool to discover planets through the gravitational effect they have on light from more distant sources. Unlike other planet detection methods, gravitational lensing does not rely on the detection of photons from the planet or its host star. Therefore, this method allows us to find planets well beyond the Solar neighborhood. Moreover, in contrast with the other detection methods, gravitational microlensing is best suited to detecting planets beyond their stars' snowline \citep[for a review, see][]{2012ARA&A..50..411G, 2018Geosc...8..365T}.

In most microlensing events, the primary observable is the Einstein timescale:
\begin{equation}
    t_E = \frac{\theta _E}{\mu _{rel}},
\end{equation}
\noindent where $\mu_{rel}$ is the lens-source relative proper motion and $\theta_E$ is the angular Einstein radius, defined as
\begin{equation}
    \theta _E = \sqrt{\kappa M_L \pi _{rel}},
\end{equation}
\noindent where $M_L$ is the mass of the lens, $\pi _{rel}$ is the source-lens relative parallax and $\kappa$ is a constant. The latter quantities are defined as 
\begin{equation}
    \pi _{rel} = \left(\frac{D_L - D_S}{D_L D_S}\right) \text{AU}
\end{equation}
\noindent and
\begin{equation}
    \kappa = \frac{4G}{c^2\text{AU}} \simeq 8.14 \frac{\text{mas}}{M_\odot},
\end{equation}
\noindent where $D_S$ and $D_L$ are the distance to the source and the lens, respectively \citep{2015ApJ...802...76Y}. 

Hence, the lens's mass and distance, $M_L$ and $D_L$, and the relative motion between the source and the lens, $\mu _{rel}$, are encoded in the primary observable, $t_E$, and are difficult to disentangle. In most planetary microlensing lightcurves, it is possible to determine $\theta _E$ via the finite source effect \citep{2004ApJ...616.1204Y}. Assuming the distance to the source is known, e.g., for observations towards the Bulge, most sources are Bulge stars, the only degeneracy remaining is between the mass of the lens and the distance to the lens.

One way of breaking this degeneracy is by measuring a quantity known as the microlens parallax vector, $\boldsymbol \pi _E$, defined as
\begin{equation}
    \boldsymbol\pi _E = \frac{\pi _{rel}}{\theta _E} \frac{\boldsymbol \mu_{rel}}{\mu _{rel}}.
\end{equation}
\noindent One can measure the microlens parallax by simultaneously observing an event from two well-separated observatories. The two observatories will see a different alignment between the lens and the source, so the projected separation and time of closest alignment will be different. This requires that the two observatories are far enough from each other, $\mathcal{O}(\text{1 AU})$, for the lightcurves to be significantly different \citep{10.1093/mnras/134.3.315, 1994ApJ...421L..75G}. 

At $>$1 AU away from Earth for the past 6 years, the \textit{Spitzer} Space Telescope \citep{2004ApJS..154....1W} was ideal for measuring microlens parallax (PI: A. Gould; PID 10036, 11006, 12013, 12015, 13005, 14012, PI: S. Dong; PID: 13250, PI: S. Carey; PID 14121). Following the successful 2014 pilot program, \textit{Spitzer} took on a new role as a ``microlens parallax satellite" with the primary objective of measuring the galactic distribution of exoplanets towards the bulge \citep{2015ApJ...799..237U, 2018ApJ...853...70U, 2015ApJ...802...76Y, 2015ApJ...810..155Y, 2015ApJ...804...20C, 2015ApJ...814...92C, 2018AJ....155..261C, 2019AJ....157..121C, 2015ApJ...805....8Z, 2015ApJ...814..129Z, 2016ApJ...825...60Z, 2017AJ....154..210Z, 2017ApJ...849L..31Z, 2015ApJ...814..111S, 2016ApJ...831..183S, 2017ApJ...840L...3S, 2019AJ....157..106S, 2016ApJ...819...93S, 2016ApJ...823...63P, 2016ApJ...820...79B, 2016ApJ...828...53H, 2017ApJ...834...82H, 2018ApJ...859...82H, 2017ApJ...838..154C, 2019ApJ...871..179C, 2017AJ....154..176S, 2018ApJ...863...23S, 2018AJ....155...40R, 2018ApJ...858..107A, 2018ApJ...860...25W, 2019ApJ...873...30S, 2019AJ....158...28J, 2019MNRAS.488.3308L, 2019arXiv191200038Z, 2020ApJ...891....3Z, 2020JKAS...53....9G, 2020arXiv200409067H}. 

\subsection{Decorrelation Techniques Developed for Spitzer Observations}
We need accurate lens properties, so it is crucial to obtain excellent photometry. However, extracting high precision photometry from \textit{Spitzer} observations can be challenging. The channel 1 (3.6 $\mu$m) of the IRAC instrument \citep{2004ApJS..154...10F} was used for the \textit{Spitzer} Microlensing Campaign. With a mean pixel scale of 1.221"/pixel and a point spread function (PSF) with a mean full width at half maximum (FWHM) of 1.66", the images are moderately undersampled. The systematics are primarily due to the convolution of changes in the telescope pointing and the significant variation of the sensitivity across each pixel. Inconveniently, the instrumental systematics can mascarade as the signature of planetary companions \citep[e.g.,][]{2016ApJ...823...63P}. 

Many techniques for decorrelating structured noise from astrophysical signals in IRAC data have been developed over the past decade. \cite{2016AJ....152...44I} tested the effectiveness of widely used decorrelation methods. They reported that Pixel Level Decorrelation (PLD) \citep{2015ApJ...805..132D}, BiLinearly Interpolated Subpixel Sensitivity mapping \citep{2012ApJ...754..136S} and Independent Component Analysis \citep{2014ApJ...786...22M, 2015ApJ...808...56M} perform best to recover the underlying astrophysical signal.

The \textit{Spitzer} microlensing team currently uses a Point Response Function (PRF) method -- which contain information about the PSF, the detector sampling and the intra-pixel sensitivity variation \citep{2015ApJ...814...92C}. The method is tailored to work on time series data and, additionally, allows multiple sources to be fitted simultaneously. %\textbf{
The current photometric extraction pipeline works for most targets, but there are cases showing possible residual red noise, e.g. OGLE-2016-BLG-1195Lb \citep{2017ApJ...840L...3S} and KMT-2018-BLG-0029 \citep{2020JKAS...53....9G}. In particular, the single lens model fitted to the \textit{Spitzer} observations of OGLE-2015-BLG-0448 resulted in residuals with significant trends as seen in Figure 2 of \cite{2016ApJ...823...63P} and reproduced in the bottom-right panel of our Figure \ref{fig: ob150448_fit}.%}
\cite{2016ApJ...823...63P} acknowledged that the residuals could be due either to a planetary companion or instrumental systematics.

Photometry extraction procedures designed for planetary transit observations have been adapted for a microlensing survey with \textit{K2} campaign 9 \citep[\textit{K2C9},][]{2016PASP..128l4401H}. \cite{2017PASP..129j4501Z} presented a photometry extraction method based on the protocol developed by \cite{2015MNRAS.454.4159H} for less crowded fields. In this method, the instrumental systematics are decorrelated against the spacecraft's pointing and are fitted simultaneously with the astrophysical microlensing model. 

\cite{2019A&A...627A..54P} introduced an open-source alternative \textit{K2} photometry extraction method build upon \cite{2016PASP..128i4503W}'s Causal Pixel Model, a data-driven instrumental model that was also developed for obtaining photometry for planetary occultations in less crowded fields. This Modified Causal Pixel Model differs in the use of the PRF to account for contamination from nearby sources.

Photometry extraction for \textit{Spitzer} is not as challenging as for \textit{K2}, as it benefits from more precise telescope pointing and smaller pixel scale, so there has been no attempt to use a photometry extraction method other than the pipeline described in \cite{2015ApJ...814...92C}. %\textbf{
However, \cite{2019arXiv190505794K} claimed that the distribution of $t_E$ and  $\boldsymbol \pi _E$ for the sample of 50 single lens events from the 2015 \textit{Spitzer} campaign cannot be reproduced by Galactic models. The authors suggested that investigating instrumental noise in the \textit{Spitzer} photometry itself may resolve the discrepancy. Meanwhile, \cite{2019ApJ...873...30S} also carried out Galactic model tests using a smaller sample of 13 published events and found the model predictions to be consistent with the \textit{Spitzer} parallaxes.%}

In similar fashion to the work of \cite{2017PASP..129j4501Z} and \cite{2019A&A...627A..54P}, we extend Pixel Level Decorrelation \citep{2015ApJ...805..132D}, developed for \textit{Spitzer} secondary eclipse observations, to microlensing campaigns. In section 2, we discuss the differences between a typical observing sequence for short-period planets and for a microlensing event. In section 3, we describe our decorrelation method. We apply our method on a few sample data sets in section 4. We summarize and conclude in section 5. 

\section{Observations}

\subsection{Challenges with Spitzer Microlensing Campaign} \label{sec:style}

Before we attempt to apply Pixel Level Decorrelation to the \textit{Spitzer} Microlensing Survey, it is important to understand the differences between typical observations of transiting planets and of a microlensing event. 

\subsection{Timescale}
Both science cases require time series photometry of a point source. However, the observing schemes are not quite the same. The duration of transiting planet observations is between a few hours for a transit or secondary eclipse to a couple days for full-orbit phase curves. Given the short length of the observations, time series observations of transiting exoplanets are generally collected in starring-mode. On the other hand, a typical microlensing event lasts for weeks which is not ideal for starring-mode observations  over the full length of the event. 

Moreover, a microlensing campaign only lasts for 40 consecutive days -- the days during which the galactic bulge is visible from \textit{Spitzer}. As the science objective of the \textit{Spitzer} Microlensing Campaign is to constrain exoplanet demographics, it is crucial to maximize the number of targets during this visibility window and to get a large temporal coverage of the microlensing event. For this reason, \textit{Spitzer} will only point at each target once or twice per day. In contrast, transiting planets are observed at a very high cadence (e.g., an image every second).

\subsection{Pointing}
\begin{figure}
\includegraphics[width=0.95\linewidth]{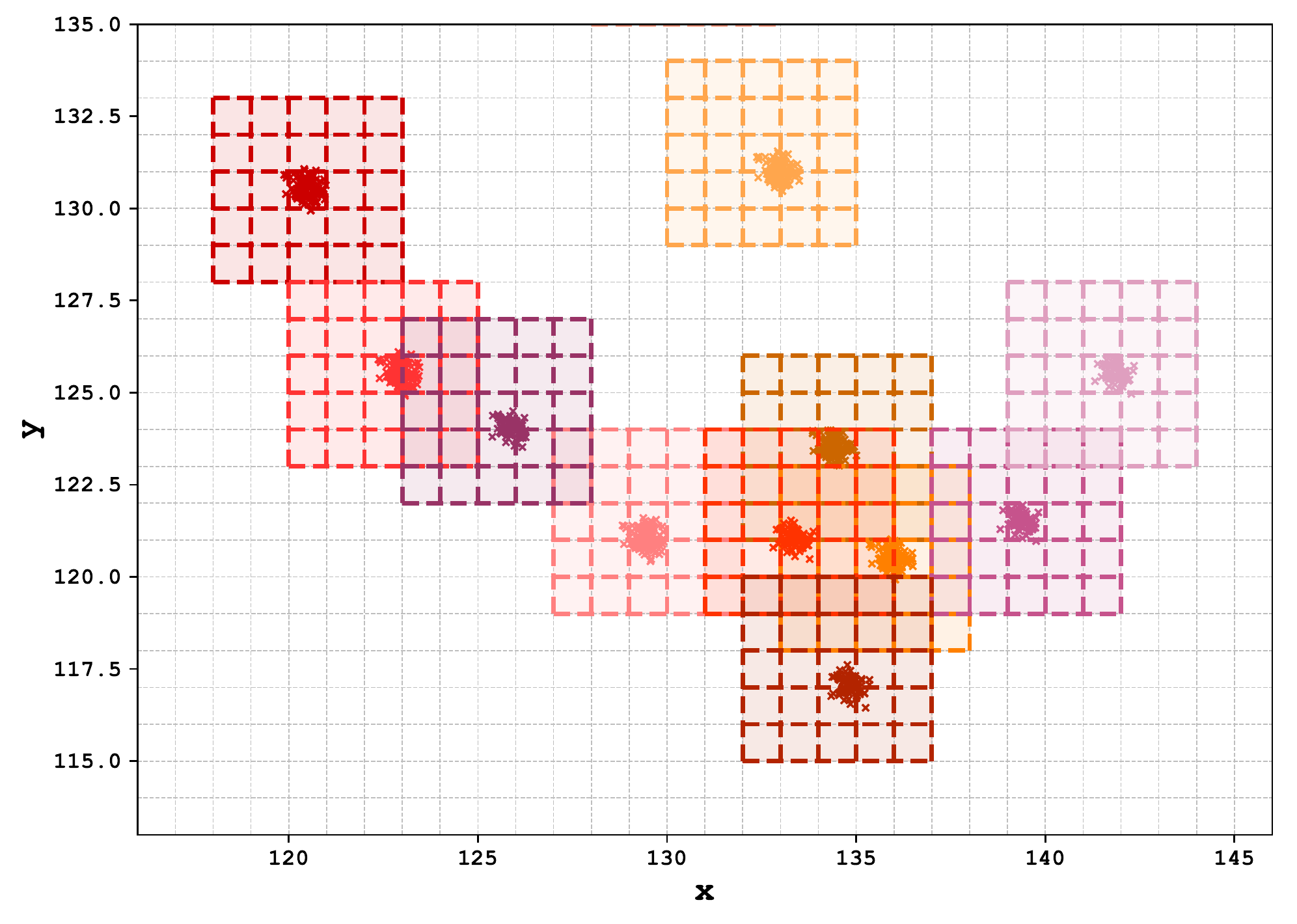}
\caption{Visualization of the position of the PSF centroids and the different dither positions on the detector for \textit{Spitzer} observations of OGLE-2015-BLG-0448. The dots represent the centroid for each observation and the different colors indicate a different dither position. The 5$\times$5 grid around the centroids at each dither position represent the pixels used to extract the target's photometry using Pixel Level Decorrelation. \label{fig:dither}}
\end{figure}

Additionally, unlike transiting planet observations, microlensing observations are dithered on six or more slightly different positions in order to minimize the contribution from bad pixels and cosmic rays and to avoid background saturation (see Figure \ref{fig:dither}). Moreover, dithering the position of the PSF on different part of the detector allows for a better characterization of the shape of the PSF. Consequently, the pointing of the telescope for a microlensing campaign is not as precise as starring-mode observations. During a starring-mode observation, the target is positioned on the ``sweet-spot'' (peak response) and the centroid of the target's point-spread function varies by $1/10$th of a pixel. On the other hand, for dithered observations towards the Galactic bulge, the centroid of the source for each dither position changes by almost a pixel along the 40-days season.

\subsection{Field Rotation}
%\textbf{
Another consequence of the long duration of microlensing events is that the orientation of the camera changes during the course of the observations. Hence, the target's PSF and that of nearby sources rotate on the detector, and this may affect the \textit{Spitzer} photometry as discussed for the observations of KMT-2018-BLG-0029 \citep{2020JKAS...53....9G}. During the 40-day microlensing season, the rotation of the field of view is about 1 degree. For staring-mode observations, on the other hand, the field rotation is negligible.%}

\subsection{Relative Brightness}
Unlike transiting systems usually observed by \textit{Spitzer}, microlensing events are often detected in crowded fields, e.g.\ towards the bulge of the galaxy. In general, transiting short-period planets observed with \textit{Spitzer} are isolated bright targets, hence, the exposure time is approximately a second to avoid saturation. Microlensing targets are fainter and the exposure time is 30 seconds.

\subsection{Magnitude of the Signal}
In general, the amplitude of transits, secondary eclipses, and phase variations are a percent or less of the stellar flux. The magnification of microlensing events depends on the source-lens projected separation but is typically up to a few magnitudes.

\subsection{Signal Coverage}
Most transiting planets observed with \textit{Spitzer} have well-known ephemerides. Hence it is possible to carefully plan the observations to cover the entire duration of the occultation. In contrast, microlensing events are unforeseeable and are first detected from ground-based surveys. These alerts then go through the selection process described by \cite{2015ApJ...810..155Y}. The ones that pass the selection process are then scheduled for \textit{Spitzer} follow-up observation. Given the unpredictable nature of microlensing events and the short visibility window of the galactic bulge, some \textit{Spitzer} microlensing targets observations do not cover the baseline and/or the peak magnification of the event.

\section{Method Description} \label{sec:floats}

Pixel Level Decorrelation (PLD) differs fundamentally from all other methods used to analyze \textit{Spitzer} data. Other decorrelation methods rely on defining an instrumental noise model that depends on the position and shape of the point-response-function (PRF). In contrast, PLD decorrelates against the intensities of the individual pixels. Unlike the current photometric pipeline used for the \textit{Spitzer} microlensing campaign, the instrumental effects are evaluated simultaneously with the microlensing model fit.

\subsection{Photometry Extraction}
For the photometric extraction, we identify the position of the target's PRF on each \textit{Spitzer}/IRAC image using the same procedure as \cite{2015ApJ...814...92C}. For each dither position, we identify the central pixel (see Figure \ref{fig:dither}), i.e.\ the pixel where the centroid of the target is most often located. Note that the centroid of the target varies by at most a pixel between epochs and the centroid of the target is sometimes located on one of the neighboring pixels. We use a 5$\times$5 pixel square aperture to obtain an initial photometric measurement for each frame. To estimate the fractional flux recorded by each pixel, we divide by the sum of intensities in the 5$\times$5 stamp. For consistency and better pixel characterization, we do not recenter the stamp on the centroid of the target for each frame, rather, we keep the stamp location fixed for each dither position (see Figure \ref{fig:dither}).

\begin{figure}
\centering
\includegraphics[width=0.60\linewidth]{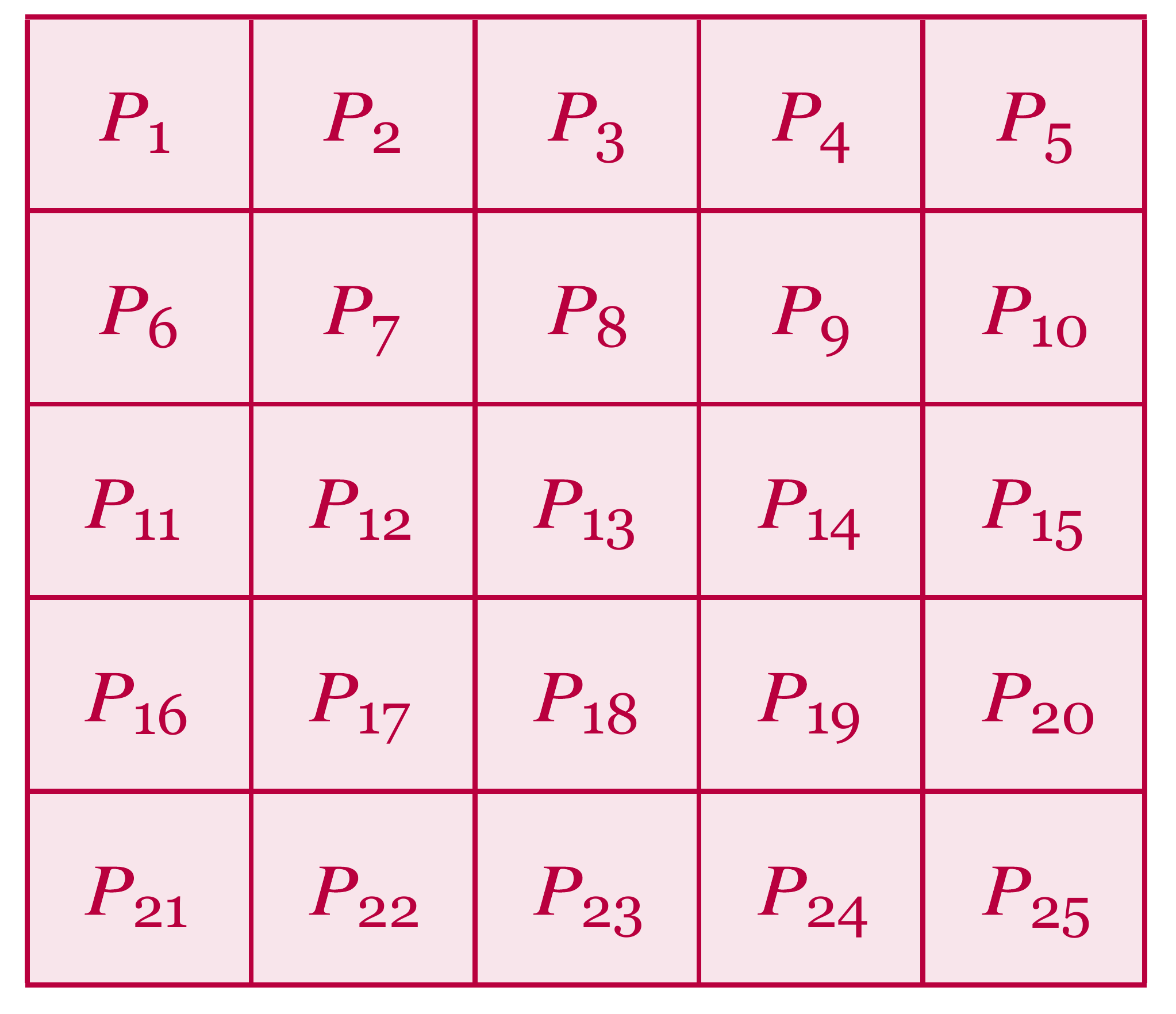}
\caption{Pixel labelling \label{fig:pixel}}
\end{figure}

The raw photometry at a given time, $F^{t_i}$, is the sum of the intensities measured by pixels within the square aperture. To characterize and remove the systematics, we define the raw photometry as a convolution of the astrophysical signal $F_{\mu lens}(t_i)$ and the instrumental noise $D(P_1^{t_i}, ..., P_{25}^{t_i})$
\begin{equation}
    F^{t_i} = \sum^{25}_{j=1} P_j^{t_i} = F_{\mu lens}(t_i) \times D(P_1^{t_i}, ..., P_{25}^{t_i}),
    \label{eq:model}
\end{equation}
\noindent where the superscript $t$ denotes the epoch of the observations and the subscript $i$ indicates the dither position, and $j$ denotes the pixel ID (see Figure \ref{fig:pixel}).

%\subsection{PLD Limitations}

\subsection{Detrending}

In PLD, the detector model is defined as some arbitrary function of the fractional flux measured by each pixel, $P_j$. To first order, we can approximate the detector model as linear:
\begin{equation}
D(P_1^{t_i}, ..., P_{25}^{t_i}) = \sum^{25}_{j=1} c_j \frac{P_j^{t_i}}{\sum _n P_n^{t_i}} = \sum^{25}_{i=1} c_j \hat{p}_j^{t_i},
\label{eq: pld coeff}
\end{equation}
\noindent where $c_j$ is the PLD coefficient corresponding to the $j$th pixel and $\hat{p}_j^{t_i}$ is the fractional flux measured by the $j$th pixel at a given time, ${t_i}$. As the instrumental systematics is not expected to behave the same way at different location on the detector, we allow each dither position, $i$, its own independent PLD coefficients.

%\textbf{
In principle, as PLD uses pixel fractional fluxes as regressors, the detector model does not need to know what causes variations in pixel fractional fluxes to model them. In particular, any effect that leads to changes in pixel fractional fluxes such as the translation and rotation of the field would be modeled by PLD.%}

\subsection{Astrophysical Model}

We can fit a point-source-point-lens model (1L1S) and a binary-lens model (2L1S) to the \textit{Spitzer} data. The single-lens magnification $A(t)$ is modelled with a standard \cite{1986ApJ...304....1P} model with the time of closest alignment between the lens and the source, $t_0$, the impact parameter, $u_0$, and the Einstein ring crossing time, $t_E$, as fitting parameters. The magnification for a 2L1S is evaluated using the \texttt{VBBinaryLensing} algorithm \citep{2010MNRAS.408.2188B, 2018MNRAS.479.5157B} described by 4 additional parameters: the planet-host projected separation in units of $\theta_E$, $s$, the planet-host mass ratio, $q$, the angular radius of the source in units of $\theta_E$, $\rho$, and the angle between the planet-host axis and the trajectory of the source, $\alpha$. The astrophysical signal as seen from the observer, $F_{\mu lens}(t_i)$, is also dependent on the baseline flux $F_b$ and the source's flux $F_s$:
\begin{equation}
F_{\mu lens}(t_i) = F_b + F_s \cdot A(t_i).
\label{eq: astrophysical}
\end{equation}

If \textit{Spitzer} observations include the baseline and the peak of the microlensing event, then we can estimate the microlensing parameters solely using \textit{Spitzer} observations. Otherwise, we also use published ground-based observations from surveys such as the Optical Gravitational Lensing Experiment \citep[OGLE;][]{2015AcA....65....1U} to further constrain the astrophysical parameters. In this scenario, we use the \texttt{MulensModel} package to evaluate the microlensing lightcurves \citep{2019A&C....26...35P}. 

Simultaneous ground-based and space-based observations allow us to constrain the microlens parallax vector components, $\boldsymbol\pi _E$ = ($\pi_{E,N}$, $\pi_{E,E}$). We use the location coordinates of the ground-based telescope and \textit{Spitzer} for the duration of the observations to obtain the magnifications as seen from each observatory, $A_g(t_i)$ and $A_s(t_i)$.  Hence, the astrophysical signal for the ground-based telescope and space-based satellite, $F_{\mu lens, g}(t_i)$ and $F_{\mu lens, s}(t_i)$ are modelled as:
\begin{equation}
\begin{split}
    F_{\mu lens, g}(t_i) & = F_{b,g} + F_{s,g} \cdot A_g(t_i) \\
    F_{\mu lens, s}(t_i) & = F_{b,s} + F_{s,s} \cdot A_s(t_i)
\end{split}
\end{equation}

\noindent where $F_{b,g}$ and $F_{b,g}$ are the baseline flux for the ground-based and space-based lightcurves, respectively, and $F_{s,g}$ and $F_{s,s}$ are the source flux as seen from the ground-based and space-based observatory, respectively.

\subsection{Regressors}
As mentioned above, we first obtained a raw time series photometry $\matr{Y}$ with $N_{dat}$ data using aperture photometry. We initially fit a 1L1S or 2L1S  model to the raw light curve to obtain initial estimates for the astrophysical parameters. Since PLD uses the fractional flux from each pixel as regressors, the total number of regressors is $N_{reg}=25\times N_{dither}$, where $N_{dither}$ is the number of dither position for a given set of observations. We can now express equation \ref{eq:model} in vector form as
\begin{equation}
    \matr{Y} = \matr{A}\matr{X}
    \label{eq: vector linear}
\end{equation}

\noindent where $\matr{A}$ is the $N_{dat} \times N_{reg}$ design matrix constructed with the set regressors $\hat{p}_j^{t_i}$ from equation \ref{eq: pld coeff} multiplied by the astrophysical model $F_{\mu lens}(t_i)$ from equation \ref{eq: astrophysical}. Hence, the elements of $\matr{A}$ are defined as $a_{i,j} = F_{lens}(t_i) \hat{p}_j^{t_i}$. $\matr{X}$ is the $N_{reg} \times 1$ vector containing the PLD coefficients. 

In other words, we construct the following matrices:
\begin{equation}
\begin{split}
    \matr{Y} & = {\begin{bmatrix} y_1 \\ y_2 \\ ... \\ y_{N_{dat}}  \end{bmatrix}} \\
    \matr{A} & = {\begin{bmatrix} a_{1,1} & a_{1,2} & ... & a_{1,N_{reg}} \\ a_{2,1} & a_{2,2} & ... & a_{2,N_{reg}}  \\ ... & ... & ... & ...\\ a_{N_{dat},1} & a_{N_{dat},2} & ... & a_{N_{dat},N_{reg}} \end{bmatrix}} \\
    \matr{X} & = {\begin{bmatrix} c_1 \\ c_2 \\ ... \\ c_{N_{reg}}  \end{bmatrix}}.\\
\end{split}
\end{equation}

For a given astrophysical model, $F_{\mu lens}(t_i)$, equation \ref{eq: vector linear} is linear. Hence, we can evaluate the PLD coefficients analytically by solving the generalized least square problem for each trial astrophysical model
\begin{equation}
    \matr{X} = \begin{bmatrix}\matr{A}^T & \matr{C}^{-1} & \matr{A} \end{bmatrix}^{-1} \begin{bmatrix}\matr{A}^T & \matr{C}^{-1} & \matr{Y} \end{bmatrix}
\end{equation}

\noindent where $\matr{C}$ is the $N_{dat} \times N_{dat}$ covariance matrix of the data
\begin{equation}
\matr{C} = \begin{bmatrix} 
\sigma ^2 _{y_{1}} & 0 & ... & 0 \\
0 & \sigma ^2 _{y_{2}} & ... & 0 \\
\vdots & \vdots & \ddots & \vdots \\
0 & 0 & ... & \sigma ^2 _{y_{N_{dat}}}
\end{bmatrix}.\\
\end{equation}

\subsection{Fitting Process} \label{sec:fitting process}

To fit the lightcurves, we develop two fitting strategies. The first is to fit only the \textit{Spitzer} data and the second is to simultaneously fit ground-based and space-based observations when the \textit{Spitzer} data alone are insufficient to constrain the microlensing parameters. In either case, we use a Markov Chain Monte Carlo (MCMC) to estimate the fit parameters and their uncertainties.

\subsubsection{\textit{Spitzer} Observations Solely}
To demonstrate our new detrending approach, we re-analyzed published \textit{Spitzer} microlensing data \citep{2016ApJ...823...63P, 2018AJ....155..261C}. We first fit the astrophysical model to the raw photometry with a Levenberg--Marquardt (L-M) algorithm, using parameter values close to those from the literature as initial values. 

The estimates obtained from the initial minimization are then used as initial guesses for our MCMC. We use \texttt{emcee} \citep{2013PASP..125..306F} to estimate the parameters that maximize the log-likelihood, $\ln L$:
\begin{equation}
\begin{split}
   \ln L = & -\frac{\chi_{Spit} ^2}{2}- \sum _{i=1} ^{N_{dat}} \ln \sigma _{y_i}  \\
   & - \frac{N_{dat}}{2} \ln  2\pi + \ln L_{reg}
\end{split}
\end{equation}
\noindent where $\chi_{Spit} ^2$ is the usual badness-of-fit,
\begin{equation}
   \chi ^2 _{Spit}= \sum ^{N_{dat}} _{i=1} \frac{[F(t_i) - F_{\mu lens}(t_i)\times D(P^{t_i}_1, ...)]^2}{\sigma^2 _{y_i}},
\end{equation}
\noindent and $\ln L_{reg}$ is an added term to constrain the flexibility of PLD and is defined as
\begin{equation}
\begin{split}
   \ln L_{reg} = & -  \sum ^{N_{dat}} _{i=1} \frac{[F(t_i) - F_{\mu lens}(t_i)]^2}{2 \sigma ^2 _{raw}}\\
   &  - N_{dat}\ln \sigma _{raw}  - \frac{N_{dat}}{2} \ln  2\pi
\end{split}
\end{equation}

\noindent where $\sigma _{raw}$ is an estimate of the photometric scatter of the raw data. Without this term, our detector model tends to overfit the data and absorb the astrophysical flux variation. Since we are confident that the slow and large flux variation is due to microlensing magnification of the primary lens, we use the difference between the astrophysical model and raw lightcurve to regulate our MCMC.

We first fit the astrophysical model to the raw photometry with a L--M and use this fit as the initial guess for our MCMC. We initialize 300 MCMC walkers widely distributed around this guess. Note that only the astrophysical parameters are jump parameters: the PLD coefficients are evaluated analytically at each step of the MCMC. We perform an initial burn-in to let the walkers explore a wide region in parameter space during which each walker performs 300 steps to identify the region in parameter space that yields the greatest log-likelihood. We then perform an MCMC where the walkers are spread around the best parameter space region until our MCMC walkers converge. To insure the convergence of our MCMC, we require that 1) the log-likelihood of the best fit does not change over last 1000 steps of the MCMC chain and 2) the distribution of walkers along each parameters over the last 1000 steps is constant. Lastly, we build a posterior probability distribution and compute the 1$\sigma$ confidence region of each parameter by marginalizing over all the walkers, over the last 1000 MCMC steps, along each parameter.

\subsubsection{Simultaneous Ground-Based and \textit{Spitzer} Observations}

If we cannot evaluate the microlensing parameters from the \textit{Spitzer} observations alone, we use published ground-based observations to further constrain the microlensing parameters by measuring the microlens parallax. As explained in section 3.3, both ground-based and space-based lightcurves are generated and fitted simultaneously. To accommodate the additional data set, we modify the log-likelihood function to
\begin{equation}
\begin{split}
   \ln L = & -\frac{\chi_{Spit} ^2}{2}- \sum _{i=1} ^{N_{dat}, s} \ln \lambda _{s} \sigma _{y_i, s}  \\
    & - \frac{N_{dat, s}}{2} \ln  2\pi + \ln L_{reg} + \ln L_{ground},
\end{split}
\end{equation}
\noindent where $\ln L_{ground}$ accounts for the goodness-of-fit to the ground-based observations and is defined as 
\begin{equation}
\begin{split}
   \ln L_{ground} = & -  \sum ^{N_{dat, g}} _{i=1} \frac{[F_{g}(t_i) - F_{\mu lens, g}(t_i)]^2}{2 (\lambda _{g} \sigma  _{{y_i, g}})^2}\\
   & - \sum _{i=1} ^{N_{dat, g}} \ln \lambda _{g} \sigma _{y_i, g}  - \frac{N_{dat, g}}{2} \ln  2\pi .
\end{split}
\end{equation}
\begin{figure*}
\includegraphics[width=.495\linewidth]{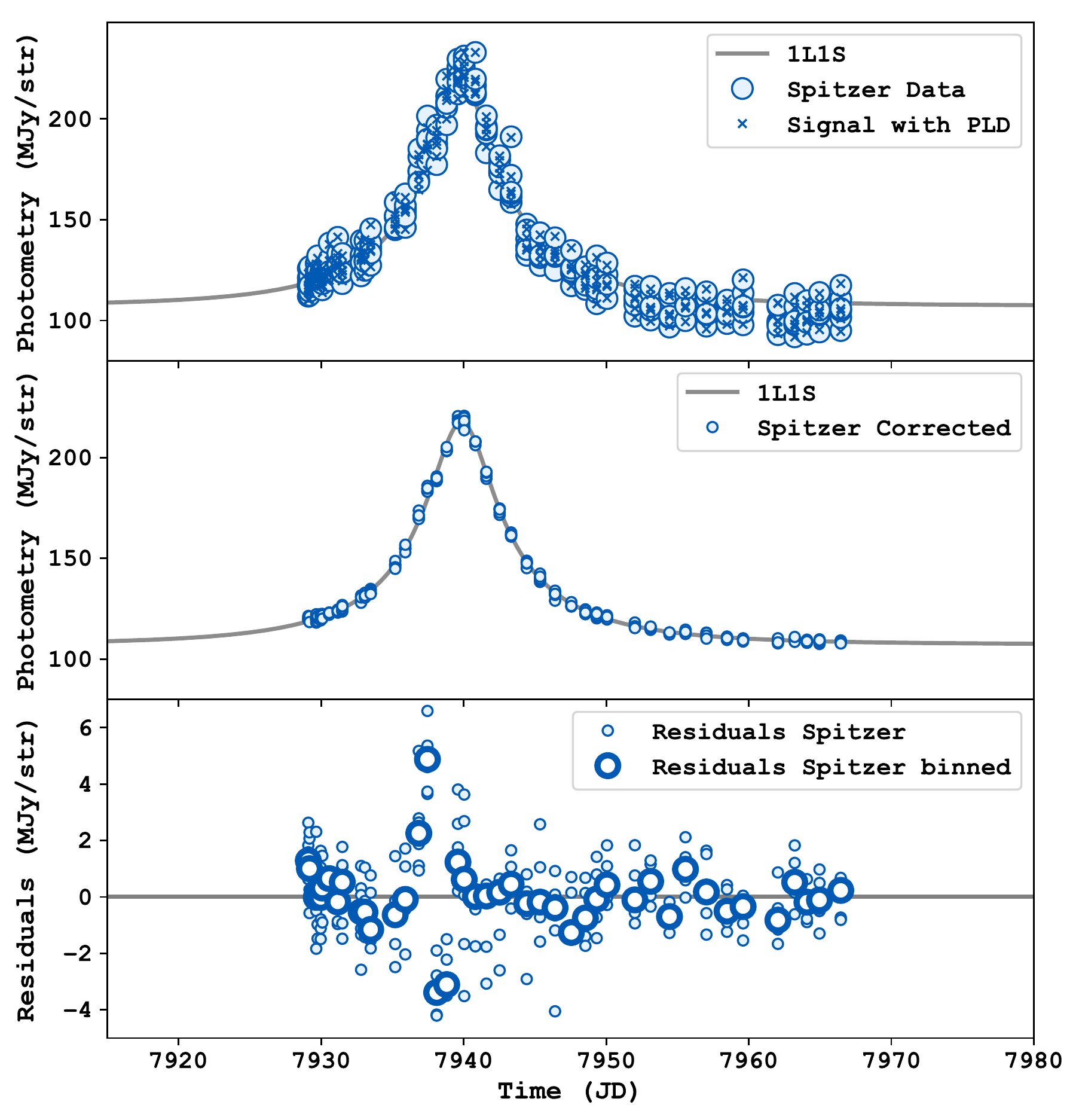}
\includegraphics[width=.495\linewidth]{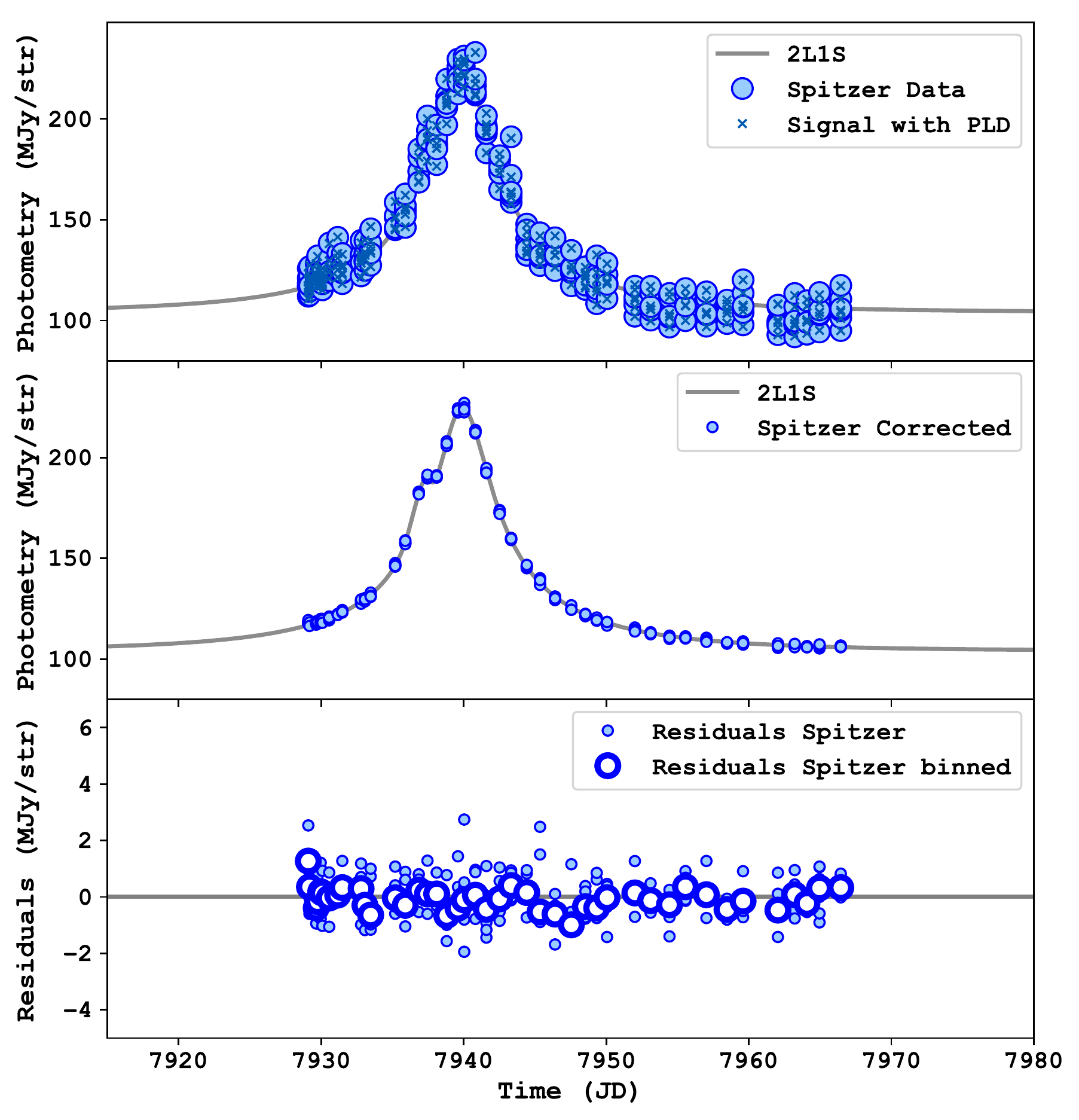}
\caption{The single-lens-single-source (1L1S; left) and binary-lens-single-source (2L1S, right) fits to the \textit{Spitzer} observations of OGLE-2017-BLG-1140. The top panels show the raw aperture as the blue circles, while the blue x's represent the best-fit model including the detector noise model. The grey line represents the best-fit astrophysical model. The second panels show the corrected photometry after Pixel Level Decorrelation. Again, the grey line represents the best-fit astrophysical model. The bottom panels show the residuals as filled blue circles and binned residuals for each epoch as white-filled circles.\label{fig: ob171140_fit}}
\end{figure*}

The $s$ and $g$ subscripts denote the space-based and ground observations, respectively. Following \cite{2012ApJ...755..102Y} and \cite{2018AJ....155..261C}, we multiply the \textit{Spitzer} uncertainties, $\sigma _{y_i, s}$ and $\sigma _{raw}$, by a factor, $\lambda _{s}$, and the ground-based observations uncertainties, $\sigma _{y_i, g}$, by $\lambda _{g}$. We then fit the data using the procedure in the previous subsection with the addition of the scaling factors as fit parameters.

\subsection{Model Comparison}
%\textbf{
While most microlensing analyses use $\chi ^2$ as a metric for model comparison, we opted to use the Bayesian Information Criterion (BIC) instead, particularly when comparing models with different number of parameters such as the 1L1S and 2L1S models. Generally, increasing the number of parameters in a model also increased its flexibility. Consequently, the fit using a model with more parameters is likely to have a smaller $\chi ^2$. For this reason, we compute the BIC for each model \citep{schwarz1978estimating, wit2012all}:
\begin{equation}
    BIC = N_{\rm par} \ln  N_{\rm dat} - 2 \ln L
\end{equation}
\noindent where $N_{\rm par}$ is the number of fit parameters, $N_{\rm dat}$ is the total number of ground-based and space-based data points and $\ln L$ is the log-likelihood function defined in section \ref{sec:fitting process}.%}

\section{Examples}
We applied our method to two microlensing events. Note that, we assume that the target is the only time-variable source in the aperture box used to extract the raw photometry. In principle, if the nearby sources are stable, then the time-varying contamination from these sources will be modelled by PLD. However, if they are variable sources, then one would need to modeled their variability as well, since PLD only decorrelates against non-astrophysical variation. For this reason, we selected targets that are located in less crowded fields or with only stable and fainter sources nearby.

The first OGLE-2017-BLG-1140, has \textit{Spitzer} observations covering the entire event, allowing us to constrain the microlensing parameters solely with the space-based data. For the second example, OGLE-2015-BLG-0448, the \textit{Spitzer} data do not cover the entire microlensing event. We therefore perform a simultaneous fit to the \textit{Spitzer} and OGLE observations. 

%\textbf{
Moreover, we selected these published events because the \textit{Spitzer} observations offer good coverage of the entire event. Note that typical microlensing \textit{Spitzer} observations do not benefit from the same generous coverage due to observing strategy of the \textit{Spitzer} microlensing program described in \cite{2015ApJ...810..155Y}. Poor coverage of an event makes the estimation of the microlensing parameters more challenging, in particular the microlens parallax parameters. Since our objective is to test the effectiveness of PLD, we chose two events with good coverage to adequately test the performance of the detector model during the fitting process.%}

\subsection{OGLE-2017-BLG-1140b}

The event OGLE-2017-BLG-1140, (RA, Dec) = (17:43:31.93, -24:31:21.6) was first analyzed by \cite{2018AJ....155..261C}. In their analysis, the \textit{Spitzer} lightcurve exhibits a stronger deviation from the single lens model than the deviation in ground-based lightcurve. We opt to apply our method solely to the space-based observations as it covers the entire event allowing us to estimate the microlensing parameters without ground-based observations. We first fit a single lens model to the raw lightcurve to estimate an initial root mean square (RMS) residuals of 6.80 MJy/str. We fit a single lens and binary lens models to the observations using the method described in section 3.5.1; the results are listed in Table \ref{tab:ob171140} and shown in Figure \ref{fig: ob171140_fit}. While \cite{2018AJ....155..261C} do not report the best-fit parameters for their 1L1S fit, their best-fit parameters for the binary lens model to the \textit{Spitzer} lightcurve are within $3 \sigma$ of ours. 

The final residual RMS of our PLD method is $\sim 4$ and $\sim 9$ times lower than the raw photometric scatter, for the single lens and binary lens fit, respectively. We report each fit's Bayesian Information Criterion (BIC) to compare their goodness of fit. There is a noticeably larger scatter in the residuals from the single lens model in Figure \ref{fig: ob171140_fit}. The scatter is significantly more pronounced near the time of planetary anomaly predicted by the binary lens model. Additionally, we see in Table \ref{tab:ob171140} that the binary lens model is strongly preferred with a $\Delta$BIC$>1000$. This test confirms that our detector model is not overfitting the planetary anomaly in the \textit{Spitzer} data.

\begin{figure}
\includegraphics[width=\linewidth]{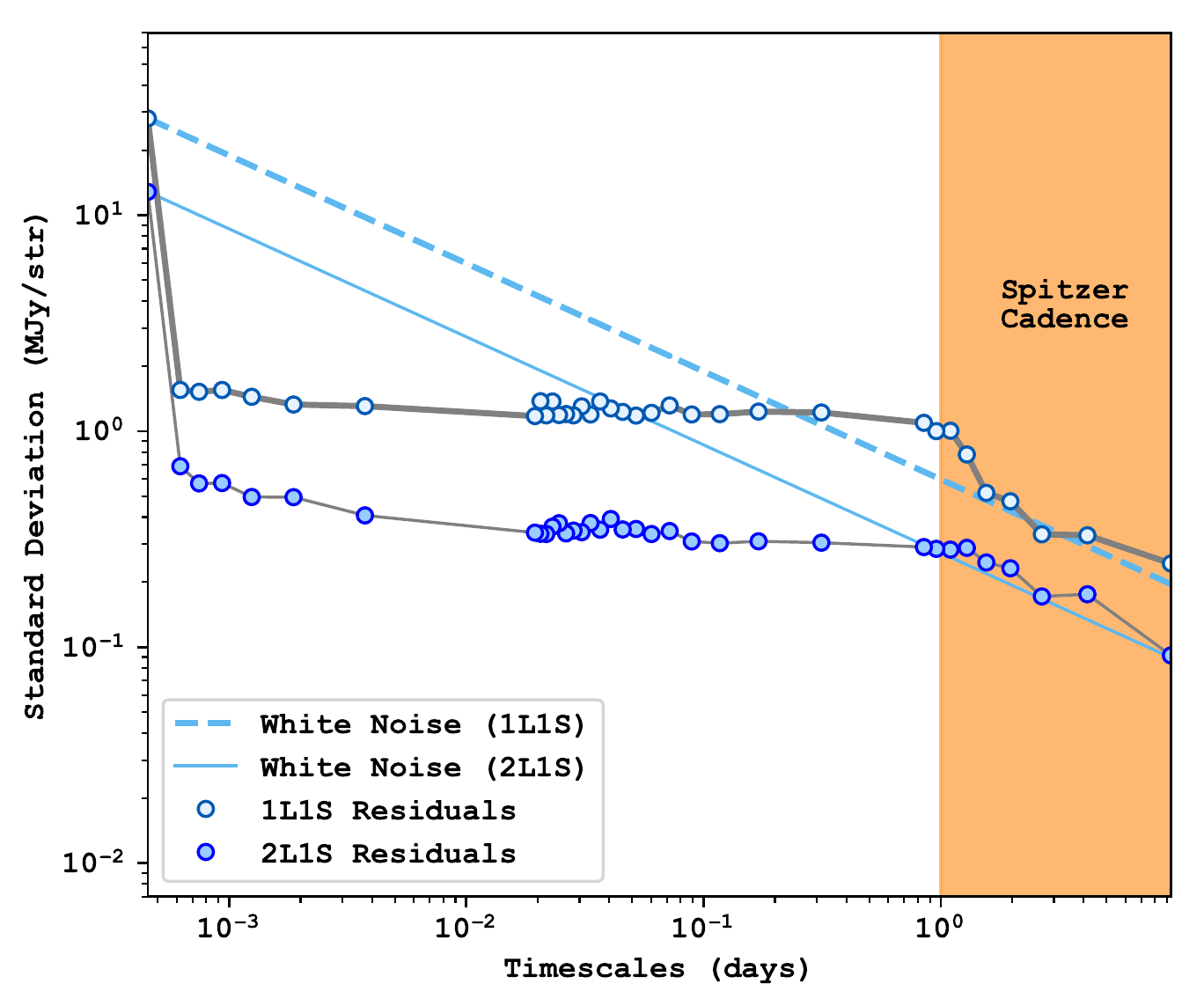}
\caption{Standard deviation of our models fit to the \textit{Spitzer} observations of OGLE-2017-BLG-1140 on different timescales. The dashed and solid light blue lines represent the expected standard deviations for the single lens and binary lens model if our residuals had been white noise. The light and dark blue dots are the calculate standard deviations for the 1L1S and 2L1S models, respectively. The orange-shaded area represents the timescales of interest for microlensing anomalies in the \textit{Spitzer} data. \label{fig:ob171140_Allan}}
\end{figure}

\begin{table}
\centering
\caption{OGLE-2017-BLG-1140 Fit Parameters Based on \textit{Spitzer} Data}
\label{tab:ob171140}
\begin{tabular}{lcc} % four columns, alignment for each
\hline
Parameter & Single Lens & Binary Lens\\
\hline
BIC                    & 5671.78                              & 4231.42                              \\ \hline
$t_0$ [HJD-2457939.0]  & $0.848     \pm 0.007               $ & $0.789     \pm 0.007               $ \\
$u_0$                  & $0.192     \pm 0.01                $ & $0.125     \pm 0.007               $ \\
$t_E$ [days]           & $12.6      ^{+0.6     }_{-0.5     }$ & $16.2      \pm 0.8                 $ \\
$s$                    & ...                                  & $0.9       \pm 0.02                $ \\
$q$                    & ...                                  & $0.0048    ^{+0.0006  }_{-0.0005  }$ \\
$\rho$                 & ...                                  & $0.032     \pm 0.003               $ \\
$\alpha$ [rad]         & ...                                  & $0.59      ^{+0.008   }_{-0.007   }$ \\
$F_{b, Spitzer}$       & $82.0      \pm 1.0                 $ & $86.5      \pm 1.0                 $ \\
$F_{s, Spitzer}$       & $26.0      ^{+2.0     }_{-1.0     }$ & $18.0      \pm 1.0                 $ \\ \hline
RMS [MJy/str]          & 1.57                                 & 0.71                                 \\

		\hline
	\end{tabular}
\end{table}

We measure the level of correlation in the residuals of our fits by calculating the standard deviation of the binned residuals (Figure \ref{fig:ob171140_Allan}). We compare them with expected standard deviations if the residuals were white noise. We see that the scatter in the residuals for the single lens fit is consistently larger than the binary fit which confirms that the binary model is preferred. Note that the step-like residual RMS is due to the non-regular cadence of the data: the roughly constant RMS scatter in Figure \ref{fig:ob171140_Allan} for timescales shorter than a day is due to the epoch cadence of the observations and the large step at very short timescale is evaluated timescale shorter than the time interval between two exposures at two different dither positions within the same epoch. %\textbf{
Since the cadence of observations is on the order of 1 day, only anomalies on longer timescales can be confidently detected;%} 
our residuals RMS reaches the white noise limit on these timescales.   

\subsection{OGLE-2015-BLG-0448}

\begin{table*}
\centering
\caption{OGLE-2015-BLG-0448 Fits Parameters Based on \textit{Spitzer} and OGLE Data}
\label{tab: ob150448 fit}
\begin{tabular}{lcccc} 
\hline
Parameter & Single Lens & Single Lens & Single Lens & Single Lens\\
& ($+,+$) & ($+,-$) & ($-,+$) & ($-,-$)\\
\hline
BIC  & 18358.04  & 18352.85 & 18367.05 & 18358.76 \\ \hline
$t_0 [HJD-2457213.0]$                  & $0.161^{+0.009   }_{-0.01    }$ & $0.162\pm 0.009               $ & $0.162\pm 0.009               $ & $0.160\pm 0.009               $ \\
$u_0$                  & $0.0875    ^{+0.0007  }_{-0.001   }$ & $0.0881    ^{+0.0007  }_{-0.001   }$ & $-0.0879   ^{+0.001   }_{-0.0007  }$ & $-0.088    ^{+0.0009  }_{-0.0007  }$ \\
$t_E$ [days]           & $60.8      ^{+0.6     }_{-0.4     }$ & $60.6      ^{+0.6     }_{-0.4     }$ & $60.6      ^{+0.6     }_{-0.4     }$ & $60.6      ^{+0.5     }_{-0.4     }$ \\
$\pi_{E,N}$            & $-0.0178   \pm 0.0003              $ & $-0.136    \pm 0.001               $ & $0.1145    ^{+0.0008  }_{-0.0012  }$ & $0.0014    \pm 0.0003              $ \\
$\pi_{E,E}$            & $-0.0922   ^{+0.0008  }_{-0.0007  }$ & $-0.0886   ^{+0.0008  }_{-0.0006  }$ & $-0.1084   ^{+0.001   }_{-0.0008  }$ & $-0.0968   ^{+0.0009  }_{-0.0007  }$ \\
$F_{b, OGLE}$          & $0.04      ^{+0.05    }_{-0.03    }$ & $0.02      ^{+0.05    }_{-0.03    }$ & $0.03      ^{+0.05    }_{-0.03    }$ & $0.02      ^{+0.04    }_{-0.03    }$ \\
$F_{s, OGLE}$          & $4.71      ^{+0.04    }_{-0.05    }$ & $4.73      ^{+0.04    }_{-0.05    }$ & $4.73      ^{+0.04    }_{-0.05    }$ & $4.73      ^{+0.04    }_{-0.05    }$ \\
$F_{b, Spitzer}$       & $32.1      \pm 0.3                 $ & $32.0      \pm 0.3                 $ & $32.4      \pm 0.3                 $ & $32.1      \pm 0.3                 $ \\
$F_{s, Spitzer}$       & $14.3      \pm 0.1                 $ & $14.7      ^{+0.1     }_{-0.2     }$ & $13.7      \pm 0.1                 $ & $14.3      \pm 0.1                 $ \\
$\lambda _{OGLE}$        & $2.0       \pm 0.2                 $ & $2.0       \pm 0.2                 $ & $2.0       \pm 0.2                 $ & $2.0       \pm 0.2                 $ \\
$\lambda _{Spitzer}$     & $2.58      \pm 0.04                $ & $2.58      ^{+0.04    }_{-0.03    }$ & $2.59      \pm 0.04                $ & $2.58      \pm 0.04                $ \\ \hline
RMS [MJy/str]  & 1.44 & 1.43 & 1.44 & 1.43 \\ \hline
	\end{tabular}
\end{table*}

\begin{figure*}
\includegraphics[width=.495\linewidth]{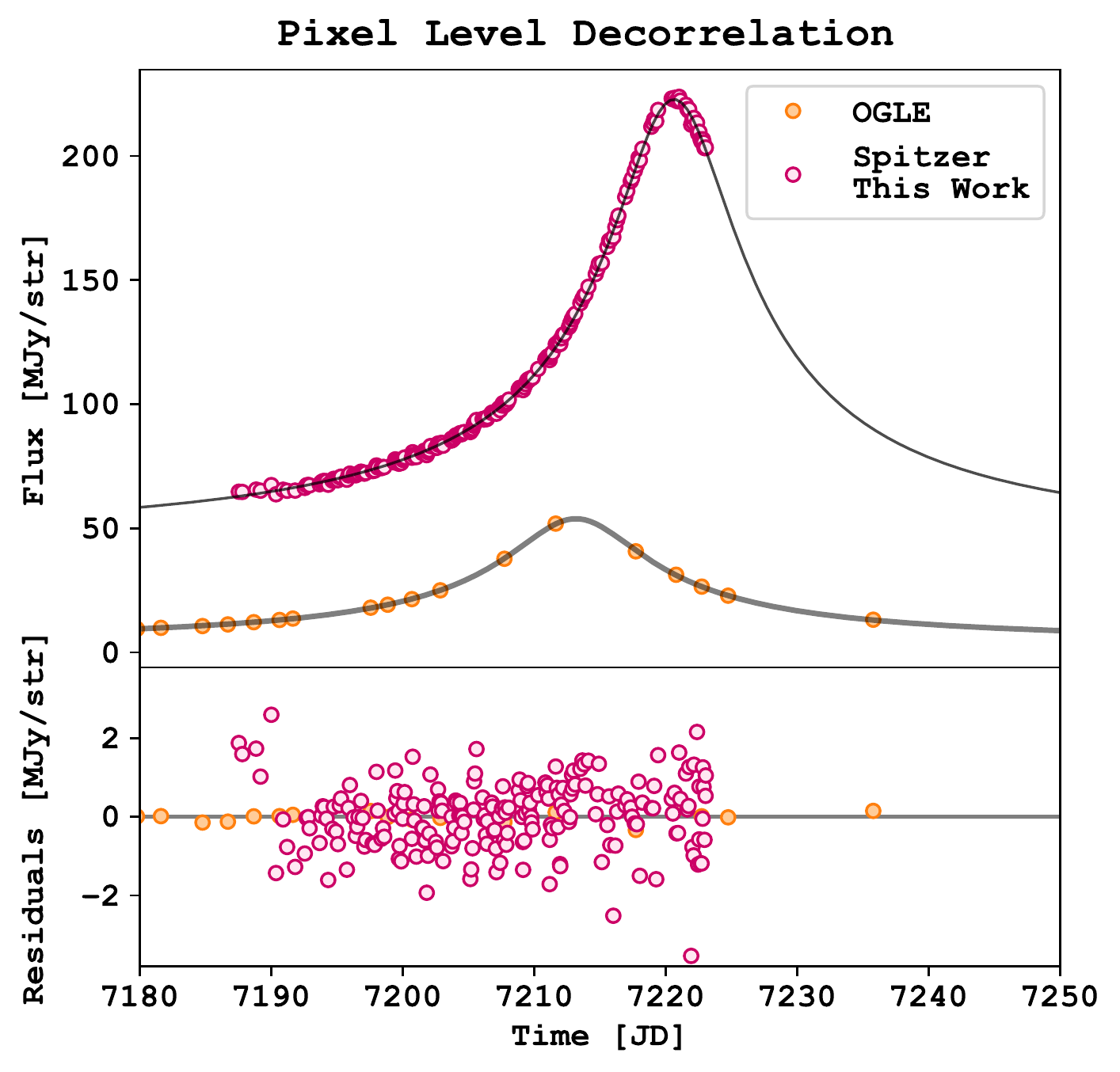}
\includegraphics[width=.495\linewidth]{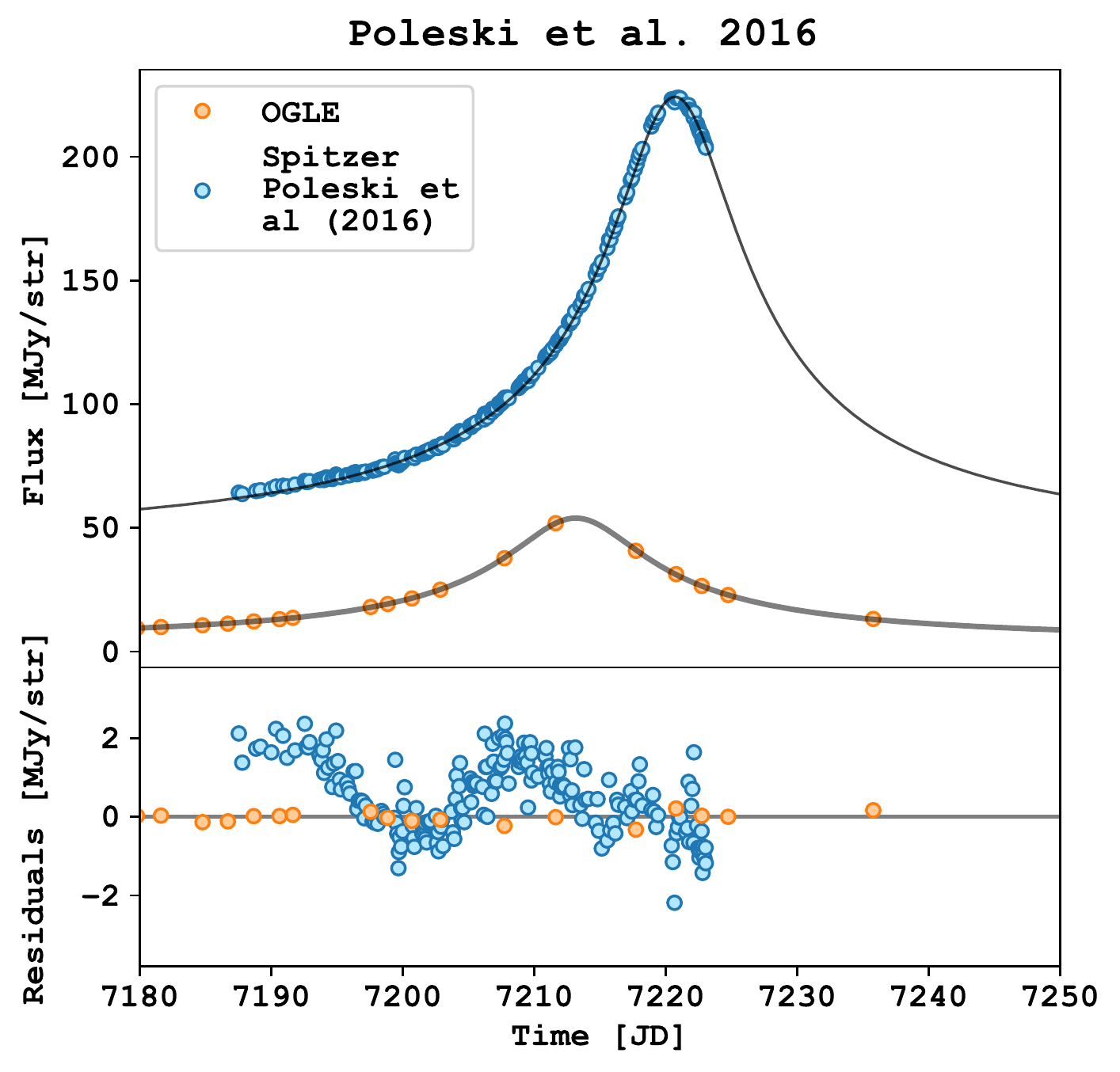}
\caption{Our best-fit 1L1S model with the lowest BIC, ($+,-$), and \textcolor{blue}{Poleski et al (2016)}'s 1L1S model with the lowest $\chi ^2$, ($-,+$), are shown in the left and right panel, respectively. In the top panels, the pink and blue dots represent our PLD corrected lightcurve and the photometry obtained on the current \textit{Spitzer} Microlensing pipeline described in \textcolor{blue}{Calchi Novati et al. (2015b)}, respectively. The orange dots represent the OGLE photometry. In the bottom panels, the pink and blue dots represent the residuals from our PLD decorrelation and the current \textit{Spitzer} microlensing pipeline, respectively. Note that PLD removes the correlated residuals in the \textcolor{blue}{Poleski et al (2016)} data that could be mistaken for a planetary anomaly.} \label{fig: ob150448_fit}
\end{figure*}

\begin{figure}
\includegraphics[width=\linewidth]{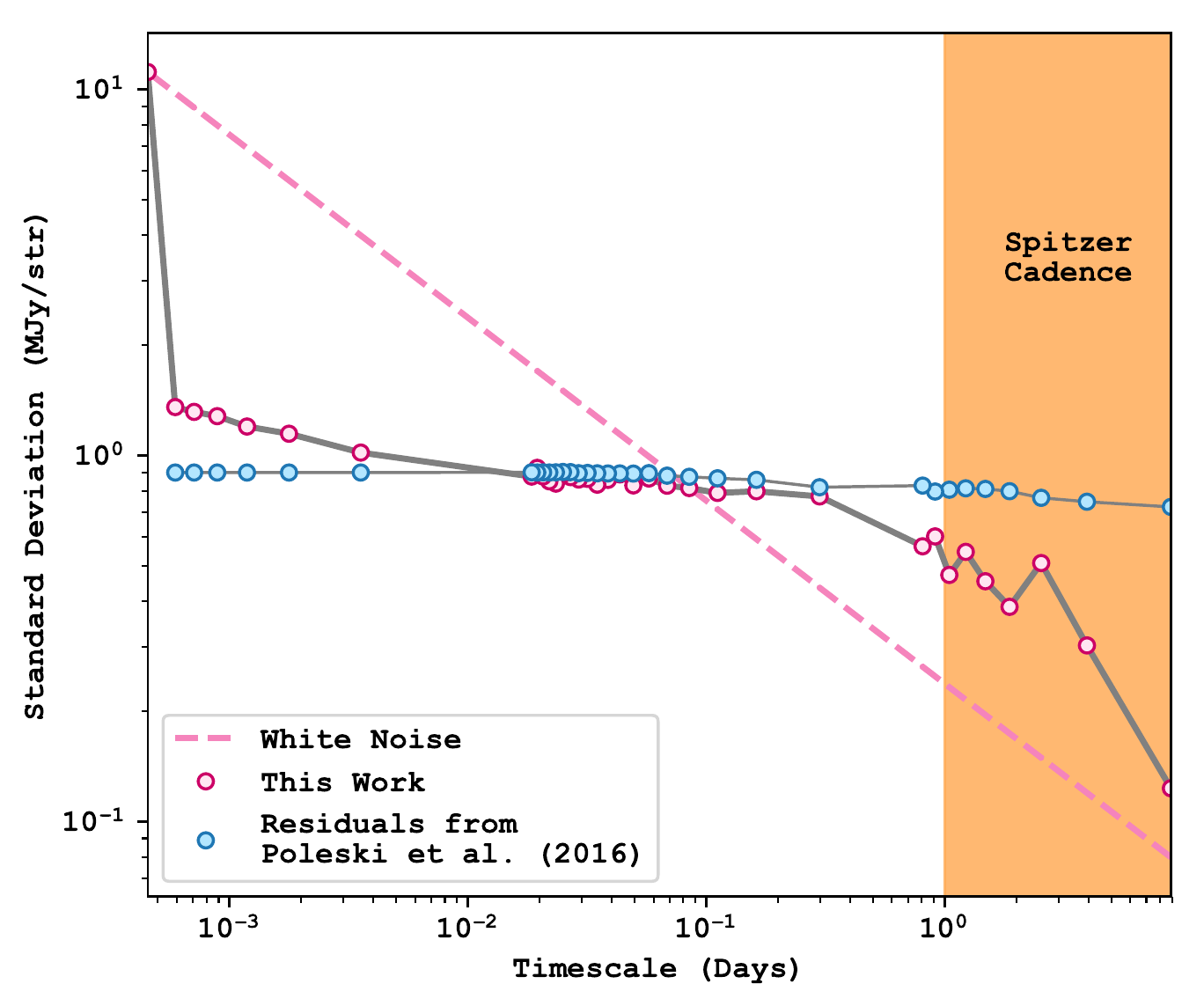}
\caption{The standard deviation of our PLD corrected 1L1S ($+,-$) residuals and \textcolor{blue}{Poleski et al (2016)}'s 1L1S ($-,+$) residuals are represented by the pink and blue dots, respectively. The dashed pink line represents the expected standard deviations if our residuals had been white noise. The orange-shaded area represents the timescales of interest for microlensing anomalies in the \textit{Spitzer} data. \label{fig:ob150448_Allan}}
\end{figure}

The event OGLE-2015-BLG-0448, (RA, Dec) = (18:10:14.38, -31:45:09.4), was first presented by \cite{2016ApJ...823...63P} who used the photometry from \cite{2015ApJ...814...92C}. The ground-based observations display no significant deviation from the single lens model. The \textit{Spitzer} residuals, however, show an obvious deviation from a 1L1S model. There are many possible explanations for correlated residuals: 1) leftover instrumental systematics or 2) possible microlensing origin. The latter is possible in principle since the ground-based and space-based observatories probe different parts of the Einstein ring \citep{2013ApJ...779L..28G}. For example, the binary anomaly of OGLE-2018-BLG-1130 was only detected by \textit{Spitzer} while showing no binarity in the ground-based observations \citep{2018ApJ...860...25W}.

For OGLE-2015-BLG-0448, \cite{2016ApJ...823...63P} explored different possible microlensing scenarios to explain the data: 1) a binary source or 2) a binary lens. The binary source hypothesis is ruled out as the source is already very red, so it would also appear in the OGLE data. As for the binary lens scenario, they attempt a binary lens fit to the observations and find that a Saturn-mass planet can explain the observations. They also note that the best-fitting binary lens model does not remove all of the time-correlated residuals, hence, unmodeled systematics is not ruled out. Consequently, they do not claim to have detected a planet.

We apply our PLD method to this event by fitting a single-lens-single-source 1L1S model to the data. Since the \textit{Spitzer} observations only partially cover the event, we used OGLE ground-base observations to further constrain the microlensing parameters as described in section 3.5.1. %\textbf{
For the OGLE data, we used 59 data points from HJD = 2457084.88043 to HJD = 2457301.5155. %}
The single lens model suffers from a four-fold degeneracy. The results from our fits are presented in Table \ref{tab: ob150448 fit}. The raw \textit{Spitzer} photometry had an RMS scatter 5.90 MJy/str and our method reduced the scatter by a factor of $\sim 4$.  

In Figure \ref{fig: ob150448_fit}, we compare our corrected photometry with the photometry from \cite{2016ApJ...823...63P} obtained using \cite{2015ApJ...814...92C}'s method. As \textit{Spitzer} flux extracted by \cite{2015ApJ...814...92C} is in arbitrary units, we rescaled their lightcurve such that the peak and the baseline is equal to our decorrelated lightcurve. We note that the significant trend observed by \cite{2016ApJ...823...63P} is not present in our single lens fit residuals. Hence, the deviation from the single models in the previous analysis of OGLE-2015-BLG-0448 were likely due to detector systematics not accounted for by the \cite{2015ApJ...814...92C} reduction. To further evaluate the performance of PLD, we evaluate the residuals RMS of our best-fit single lens model with the lowest BIC at different timescales and compare it with the residual RMS from \cite{2016ApJ...823...63P}'s single lens fit. We note that \cite{2015ApJ...814...92C}'s method combines all dithers per epoch to evaluate the photometry while we use all exposure in our decorrelation, hence, evaluating the scatter for a given timescale is essential for the comparison. The standard deviations vs. timescales calculations are presented in Figure \ref{fig:ob150448_Allan}. While \cite{2016ApJ...823...63P}'s residuals are slightly less scattered at short timescales, they are significantly more correlated at timescales longer than the \textit{Spitzer} observations cadence. On a timescale of $\sim$9 days, our method reduced the noise by a factor of 5.9, when compared to the single lens and binary lens fit, respectively. On a timescale of $\sim$4 and $\sim$2.5 days, our method improves the noise by a factor of 2.5 and 1.5, respectively, in comparison with the photometry from \cite{2015ApJ...814...92C}. While our single lens residual RMS is above the photon noise limit, it still outperforms the current pipeline used for the \textit{Spitzer} microlensing campaign.

\section{Discussion and Conclusion}

We present an alternative method to extract and reduce photometry for the \textit{Spitzer} Microlensing campaigns using Pixel Level Decorrelation, a method initially developed to decorrelate lightcurves of transiting exoplanets with \textit{Spitzer}. This method uses the fractional flux recorded by each pixel as regressors to model the systematics. PLD models the instrumental systematics and flux contribution from nearby stars. Advantages of this method includes not requiring precise centroid measurements for each exposure, not needing dithered observations, and better noise reduction.

We have tested PLD on \textit{Spitzer} observations of OGLE-2017-BLG-1140 and OGLE-2015-BLG-0448. We find that PLD is able to reduce the RMS scatter in the raw photometry by at least a factor of 4. We also find that for the event OGLE-2015-BLG-0448, our decorrelation produces photometry up to an order of magnitude more precise than the \cite{2015ApJ...814...92C} pipeline on timescales of microlensing anomalies. We note that there are significant differences between the two methods. Most published \textit{Spitzer} microlensing lightcurve analyses are done in 2 steps: first reduce the photometry, then evaluate the microlensing parameters. Similarly to photometry extraction methods developed for \textit{K2} microlensing observations \citep{2017PASP..129j4501Z, 2019A&A...627A..54P}, the PLD method fits the microlensing model and the noise model simultaneously. 

%\textbf{
We note that both events tested in this work have benefited from \textit{Spitzer} observations with good coverage over the duration of the microlensing event. For events with poor coverage, the microlensing parameters will be difficult to constrain with great precision. However, to model the detector systematics with PLD, the key is to have more data to better characterize the detector noise. Even without full coverage of the event, PLD will be able to remove the systematics if there are a large number of exposures. The microlensing parameters, however, will have larger uncertainties.%}

%\textbf{
The \textit{Spitzer} Microlensing campaign has enabled unprecedented microlens parallax measurements to build a planet distribution in the galaxy. However, tensions have been claimed between the results from this campaign and prediction from commonly used Galactic models \citep{2019arXiv190505794K} suggesting that the photometry extraction could be the source of error. %}
Alternative photometry extraction schemes such as PLD could help investigate the source of the discrepancies.

The Pixel Level Decorrelation technique for noise characterisation is not uniquely applicable to \textit{Spitzer} observations. For example, PLD has been successful when applied to \textit{K2} observations of transiting exoplanets \citep{2016AJ....152..100L, 2018AJ....156...99L}. Hence, given the versatility of this method, it could be adapted to other microlensing campaigns with other space telescopes such as the \textit{Kepler} Space Telescope for the \textit{K2C9} campaign. 

%\textbf{
The forthcoming Nancy Grace Roman Space Telescope, formerly known as the Wide Field Infrared Survey Telescope \citep[\textit{WFIRST},][]{2015arXiv150303757S}, is expected to detect thousands of exoplanets via microlensing \citep{2019ApJS..241....3P}. However, its infrared detector will share similarities with \textit{Spitzer}'s ---including intra-pixel sensitivity variations--- and therefore could benefit from decorrelation methods such as PLD.%}

\section*{Acknowledgements}

We would like to thank P. Mroz for the discussions of microlensing modelling, R. Poleski for help with \texttt{MulensModel}. We would also like to thank our referee, D. Bennett, for constructive criticism.  This work was supported in part through a Visiting Graduate Researcher Fellowship (VGRF) at Caltech's Infrared Processing and Analysis Center (Caltech/IPAC), McGill University's Graduate Mobility Award, the Technologies for Exo-Planetary Science (TEPS) International Internship program, and the Natural Sciences and Engineering Research Council of Canada (NSERC)'s Postgraduate Scholarships-Doctoral Fellowship. This work is based on archival data obtained with the Spitzer Space Telescope, which is operated by the Jet Propulsion Laboratory, California Institute of Technology under a contract with NASA. Support for this work was provided by an award issued by JPL/Caltech.
%%%%%%%%%%%%%%%%%%%%%%%%%%%%%%%%%%%%%%%%%%%%%%%%%%



\bibliographystyle{mnras}
\bibliography{Spitzer-ulens} % if your bibtex file is called example.bib

%\begin{thebibliography}{99}
%\end{thebibliography}

%%%%%%%%%%%%%%%%%%%%%%%%%%%%%%%%%%%%%%%%%%%%%%%%%%

%%%%%%%%%%%%%%%%% APPENDICES %%%%%%%%%%%%%%%%%%%%%

%\appendix


%If you want to present additional material which would interrupt the flow of the main paper, it can be placed in an Appendix which appears after the list of references.

%%%%%%%%%%%%%%%%%%%%%%%%%%%%%%%%%%%%%%%%%%%%%%%%%%

% Don't change these lines
\bsp	% typesetting comment
\label{lastpage}
\end{document}